\documentclass[reprint, amsmath,amssymb,
 aps,nofootinbib,showkeys, twocolumn]{aastex63}
 \usepackage{color}
 \usepackage{amsmath}

\definecolor{orange}{RGB}{255,128,0}
 
\shorttitle{RCs with the multiSFDM model}
\shortauthors{Sol\'is-L\'opez et al.}
\begin{document}
\title{Rotation curves with the multistate Scalar Field Dark Matter model}
\correspondingauthor{Jordi Sol\'is L\'opez}
\email{jsolis@fis.cinvestav.mx}
\author[0000-0001-5564-7133]{Jordi Sol\'is-L\'opez}
\affiliation{Departamento de F\'isica, Centro de Investigaci\'on y de Estudios Avanzados del IPN, A.P. 14-740, 07000 M\'exico D.F., M\'exico.}
\author[0000-0002-5681-7699]{Luis E. Padilla}
\affiliation{Instituto de Ciencias F\'isicas, Universidad Nacional Aut\'onoma de M\'exico, Apdo. Postal 48-3, 62251 Cuernavaca, Morelos, M\'exico.}
\affiliation{Mesoamerican Centre for Theoretical Physics,
Universidad Aut\'onoma de Chiapas, Carretera Zapata Km. 4, Real
del Bosque (Ter\'an), Tuxtla Guti\'errez 29040, Chiapas, M\'exico}
\author[0000-0002-0570-7246]{Tonatiuh Matos}
\affiliation{Departamento de F\'isica, Centro de Investigaci\'on y de Estudios Avanzados del IPN, A.P. 14-740, 07000 M\'exico D.F.,  M\'exico.}
\date{\today}

\begin{abstract}
We use the concept of co-added rotation curves of Salucci \textit{et al.} to investigate the properties of axi-symmetric multistate Scalar Field Dark Matter halos in low surface brightness galaxies and dwarf disc galaxies. We fit their rotation curves in two-state configurations and we find that all of these can be well fitted with a particle mass $\mu \sim (10^{-23} - 10^{-24})\rm{eV}/c^2$. Comparing our results with the standard cosmological model, the well-known $\Lambda$-cold dark matter, by using the Bayesian information criterion and the Akaike information criterion, we found that our two-state model seemed to be preferred. 
\end{abstract}

\section{\label{sec:intro}Introduction}

It is now well accepted that to understand how galaxies and clusters of galaxies were formed, additionally to the baryonic matter, which is responsible to contribute to the gravitational pull necessary to maintain stable all these structures in the universe, an extra element known as dark matter (DM) is necessary to introduce. It is also well known that without this DM component it is difficult to explain the observed anisotropies in the cosmic microwave background radiation, the large-scale structure formation in the universe, the galactic formation process or the gravitational lenses of distant objects, among others. In this way, today it is well established that DM is a fundamental ingredient of the cosmic inventory. 

In this direction, the standard cosmological model assumes that the DM of the universe is comprised of a non-relativistic, collisionless gas -- cold dark matter (CDM) -- and usually assumed to be weak-interacting massive particles (WIMPs) which originated as a thermal relic of the Big Bang \citep{cdm1,cdm2}. Although WIMP dark matter describes observations well at cosmological scales, it is in apparent conflict with some observations on small scales \citep[e.g. the problem of cuspy-core halo density profiles, overproduction of satellite dwarfs within the Local Group, and others, see, for example,][]{cdm5,cdm6,cdm4,cdm7,cdm3}. All of these discrepancies are based on the fact that from CDM $N$-body simulations of structure formation, the CDM clusters form halos with a universal Navarro-Frenk-White (NFW) density profile at all scales \citep{NFW}, which is proportional to $r^{-1}$ (a `cuspy' profile) at small radii, whereas it decays as $r^{-3}$ for large radii.
{Furthermore, the attempts to detect WIMPs directly or indirectly \citep{cdm8} have no successful results, and a large range of parameters thought to be detectable has not been measured. To help us solve all these issues several alternative DM models have been proposed.}

{One of the strongest candidates to substitute the standard CDM} is the scalar field dark matter (SFDM) model. This model states that DM is an ultra-light real or complex scalar field, minimally coupled to gravity, and interacting only gravitationally with baryonic matter. The main idea was originated about two decades ago by \cite{sf4,sf40,sf3,sf1,sf2,sf6,sf5} and \cite{sf7}, with some hints traced further back in \cite{sf80} and \cite{sf8}. However, it was systematically studied for first time by \cite{sf13,sf14}  \citep[for a review of SFDM, see][]{rev1,rev2,RS,rev3,niemeyer2019small}.

{Over the years the idea has been rediscovered or renamed {by various authors}, the most popular names are: SFDM \citep{sf4}, wave DM \citep{sc4}, fuzzy DM \citep{sf2}, Bose-Einstein condensate DM \citep{sf10}, and ultra-light axion DM \citep{sf12,sf11}. In this work, we use the most general name SFDM.}

{The SFDM model alleviates problems at small scales  because of the dynamical properties derived from its macroscopic-sized de Broglie wavelength. It solves the cusp/core problem in CDM as seen in several cosmological simulations of structure formation \citep{sc4,sc10,sc11,sc12,sc13,sc14} in which the SFDM halos have cored density profiles within their inner {most} regions of galactic systems. This halos have a central core (referred in the literature as solitons \citep{sc15,sc16,sc17,sc14}) and are surrounded by an envelope generated by a quantum interference pattern that is well fitted by a NFW density profile.}

{Solitons have a size of similar magnitude than the  de Broglie wavelength of individual bosons: 
\begin{equation*}
    \lambda_{dB}\propto (\mu v)^{-1},
\end{equation*}
where $v$ is the ``{average} virial velocity" of the bosons and $\mu$ its mass, that, in order to reproduce galactic cores of one kiloparsec of size is typically assumed in the range of $\mu \sim (10^{-20}-10^{-22})~\rm{eV}/c^2$.}

In different simulations, it has been found  a strong scaling correlation between the mass $M_c$ of the core and the mass $M_h$ of the whole halo given by $M_c\propto M_h^{\beta}$. \footnote{We recommend \citep{mipaper2} for the extension of this core-halo mass relation in presence of baryonic components or \citep{mipaper1} when a self-interaction between the SFDM particles is allowed.} The particular value of the $\beta$ parameter is still under debate given that different authors have obtained different results.
The value $\beta = 1/3$ was found in \cite{sc4,sc10} from their fully cosmological simulations. And $\beta = 5/9$ by \cite{sc13,moczext1} adopting more simplified scenarios on galaxy formation but with better resolution. On the other hand, there have been some other works that have tried to fix this $\beta$ parameter but have not succeeded, since, they affirm, that the results of their simulations were not consistent with a single value of $\beta$ for all their simulated galaxies. The latter is consistent with the results presented in \cite{may2021structure}, in which the authors studied the scaling relations for SFDM halos. In that work, the authors used a modified version of the GADGET code (AX-gadget) to study how the different scaling relations for cores and envelopes in the SFDM are modified once incorporating the different effects that are added once studying galaxies in a real cosmological environment. Their results showed that not all galaxies can be described with a single $\beta$ and the scaling relations reported by \cite{sc4,sc10,sc13} and \cite{moczext1} are only consistent with galaxies in some limiting cases, being only valid for the most relaxed and spherical symmetric systems.

Due to the discrepancy of $\beta$ in the core-halo mass relation, it is clear that this topic is not yet closed. In this direction, it has been a related idea that was proposed for the first time in \cite{matos2007flat} in which the gravitational co-existence of different energy eigen-states of the wave function (multistates) are responsible to describe a complete galaxy in this SFDM scenario. Recently, in \cite{Nuevo} they showed a general method to find solutions of multistate configurations, this method encompasses the spherical multistates of \cite{sph_multisates}, excited single states, l-boson stars \citep{lbosons} as well as the new axi-symmetric multistates, furthermore, they show a possible formation process of this axi-symmetric configurations by the collision of single states. Although they do not give a bound, they show that the particular solutions they consider are stable. This multistates possibility is still in its infancy as the scientific community is just beginning to study this scenario. 

Our intention in this work is to test multistate SFDM profiles with rotation curves. For this purpose, we decided to use the so called universal rotation curve (URC) method, which was introduced in \cite{spiralURCI,spiralURCII} for the case of spiral galaxies, but it has also been applied to low surface brightness (LSB) galaxies \citep{URCLSB}, dwarf disc (DD) galaxies \citep{URCDD} and LSB and DD combined \citep{salucci}. The URC is the model that fits the co-added rotation curve, which is constructed from a sample of rotation curves (RCs) with normalized radius and normalized circular velocities. The URC thus is a function with two parameters: the normalized radial coordinate and a galaxy family identifier that could be, for example, the optical velocity $v_{opt}$ (measured velocity at the optical radius), galaxy luminosity $L_B$ or absolute magnitude $M_K$. {The great advantage of using URCs is that once we have found a good mass model for the co-added RC, it is possible to recover the mass model of each galaxy within that particular family.}

The article is organized as follows: In Section \ref{sec:model} we present the multistate Scalar Field Dark Matter (multiSFDM) model, the background, the properties, and the particular configurations we will use in the paper. In Section \ref{sec:DD} we present the mass model for the dwarf disc galaxies, in Section \ref{sec:LSB} the mass model for the LSB galaxies, in Section \ref{sec:discussion} the discussion of the results, finally in Section \ref{sec:concl} we give our conclusions.
\section{The scalar field dark matter model}\label{sec:model}

We solve the equations for self-gravitating scalar fields $\Psi$ of mass $\mu$ that play the field theory version of the spinless particles coupled to Einstein's gravity in the weak field regime: the three-dimensional Gross-Pitaevskii-Poisson system, in the case where there is no self-interaction, becomes the Schrodinger-Poisson (SP) system \citep{sph_multisates}:
\begin{eqnarray*}
i\hbar \frac{\partial \Psi_{nlm}}{\partial t}&=&-\frac{\hbar^2}{2\mu}\nabla^2\Psi_{nlm}+\mu V\Psi_{nlm}, \\
\nabla^2 V&=& \frac{\mu^2 c^4}{\hbar^2} \sum_{nlm}|\Psi_{nlm}|^2 
\end{eqnarray*}
where $c$ is the speed of light and $\hbar$ is the reduced Planck constant.

If we consider stationary states $\Psi_{nlm}(t,r,\theta,\varphi)=\sqrt{\frac{4\pi G}{\tilde{\mu}^2 c^2}}e^{-iE_{nlm}t/\hbar}\Phi(r,\theta, \phi)$ it becomes
\begin{subequations} \label{eq:gpp2}
\begin{eqnarray}
&&\hat{\nabla}^2 \Phi_{nlm} - 2 (\hat{V} + \hat{E}_{nlm}) \Phi_{nlm}
= 0, \label{eq:gpp_Phi} \\
&&\hat{\nabla}^2 \hat V = \sum_{nlm}|\Phi_{nlm}|^2, \label{eq:gpp_V}
\end{eqnarray}
\end{subequations}
where $G$ is the gravitational constant, $\hat{V} \equiv V/c^2$, $\hat{E}_{nlm} \equiv \frac{E_{nlm}}{\mu c^2}$ and $\tilde{\mu}\equiv \mu c / \hbar$ has units of $\text{length}^{-1}$ and makes the coordinates and the Laplace operator dimensionless: $\hat{r} = \tilde{\mu} r$ and $\hat{\nabla}^2 = \frac{1}{\hat \mu ^2}\nabla^2$.

The SP system has the scaling property 
\begin{equation}\label{rescaling}
\left(\hat r, \Phi_{nlm}, \hat V, \hat E_{nlm}, N \right) \rightarrow
\left(\hat r / \sqrt{\lambda}, \lambda \Phi_{nlm}, \lambda \hat V, \lambda \hat{E}_{nlm}, \sqrt{\lambda}N \right)
\end{equation}
that give us two free parameters for our model, the particle mass $\mu$ and the scaling parameter $\lambda$.\footnote{{Whenever more sates are considered, extra free parameters appear, those could be, for example, the ratio between wave function amplitudes $\zeta \equiv \frac{\psi_{100}}{\psi_{nlm}}$.}} Using this $\lambda$ parameter, it is possible to construct an infinite number of solutions of the SP system once one solution is known.

In what follows we work with dimensionless variables and we will drop the $\hat{}$ symbol for simplicity.

We can consider several cases for a DM halo: a) The simplest possibility is to consider a single state, when all boson particles are in the same state $\Psi_{nlm}$, with $n$ taking only one value $1,2,...$, and also for $l$ and $m$ taking one of its possible values $l = 0,1,..,n-1$ and $m = -l, -l+1,..., l$. In this case there is only one Schrodinger equation (\ref{eq:gpp_Phi}) and only one term in the RHS of Equation (\ref{eq:gpp_V}). It happens that in the single state case only the ground state $\Psi_{100}$ is stable \citep{cooling1}.

One other possibility is b) multistates (multiSFDM), states where some particles are in the ground state and some in other excited states. The DM density in the RHS of Equation (\ref{eq:gpp_V}) is then of the form $|\Psi_{100}|^2 + |\Psi_{nlm}|^2$, $n = 2,3,...$; $l = 0,1,..,n-1$; $m = -l, -l+1,..., l$, and there is one Schr\"odinger equation (\ref{eq:gpp_Phi}) for each state.  The idea is that a galaxy should be described with a collection of states. The particular value of the $n,l,m$ parameters should depend on the process of evolution and formation of the galaxy we are interested to model, so in general these parameters should not be able to be set in a general way for all types of galaxies. However, as an example and with the intention of showing the enormous advantages that these multistate configurations give us, in this work we will adopt working with scenarios of only two states, that is, we will take the ground state together with one excited state of the previous system. \footnote{The idea of using in all cases the ground state is because it has been demonstrated that for multistate configurations to be stable, the ground state must be presented in the system \citep{sph_multisates}.}  Particularly, we shall only concentrate on the multistate configurations that we present in what follows.

\subsection{multiSFDM case (100, 21m)}

Following the general framework of \cite{Nuevo} for the multiSFDM case $(\Psi_{100},\Psi_{21m})$, the system (\ref{eq:gpp2}) becomes
\begin{eqnarray} \label{eq:mix21m}
\nabla^2_{r_0} \psi_{100}( r) &=& 2 (V_{00} - E_{100})\psi_{100}, \nonumber\\
\nabla^2_{r_1} \psi_{21m}(r) &=& 2 (V_{00} + C  r^2 V_{20}-  E_{21m})\psi_{21m}, \nonumber\\
\nabla^2_{r_0} V_{00}(r) &=& \psi_{100}^2 + r^2 \psi_{21m}^2, \nonumber\\
\nabla^2_{r_2} V_{20}(r) &=& |C|\psi_{21m}^2, 
\end{eqnarray}
where we have expanded the gravitational potential in spherical harmonics $Y_{lm}(\theta, \phi)$ as 
\begin{equation*}\label{eq:pot210}
V (r, \theta) = \sqrt{4\pi} \left( V_{00}(r) %
Y_{00}(\theta, \phi) + V_{20}(r) r^2 Y_{20}(\theta, \phi) \right)
\end{equation*}
and the scalar field states have been written as $\Phi_{nlm}= \psi_{nlm}(r) r^l Y_{lm}(\theta, \phi)$. The constant $C= 2/\sqrt{5}$ for $m=0$ and $C= -1/\sqrt{5}$ for $m=\pm 1$. The $l$-laplacian operator is defined as

\begin{equation*}
\nabla^2_{r_l} \equiv \frac{\partial^2}{\partial r^2} + \frac{2(l+1)}{r} \frac{\partial}{\partial r}. 
\end{equation*}

The enclosed mass at radius r of the DM halo is
\begin{equation*}
M =\frac{c^2}{G\tilde{\mu}} N
\end{equation*}
with $N = \sum_{n,l,m} N_{nlm}$ {the dimensionless enclosed mass.} Here, the number of particles $N_{nlm}$ of each state is
\begin{equation*}
N_{nlm} = \int |\Phi_{nlm}|^2 r^2 dr d\Omega.
\end{equation*}

 The circular velocity of a particle due to this SFDM halo is given by
\begin{equation}\label{eq:halo}
 v_h^2 = \frac{P_0}{r} - \frac{\sqrt{5}}{2} r^2\left(r P_2 + 2 V_{20} \right)  
\end{equation}
where 
\begin{equation*}
    P_0 = r^2 \frac{dV_{00}}{dr}, \ P_2 = \frac{dV_{20}}{dr}.
\end{equation*}

The system (\ref{eq:mix21m}) with {the following} boundary conditions 
\begin{gather}
\begin{aligned}
\psi_{100}(r_f) &=0, \ & \frac{d\psi_{100}}{dr}\bigg\rvert_{r=0}&=0, \nonumber\\
\psi_{21m}(r_f)&=0, \ &\frac{d\psi_{21m}}{dr}\bigg\rvert_{r=0}&=0,\nonumber\\
V_{00}(r_f)&= - \frac{N_T}{r_f}, \ &P_0(r_f)&= N_T, \nonumber\\
V_{20}(r_f)&= 0, \ &P_2(0)&=0, \nonumber
\end{aligned}
\end{gather}
becomes a boundary value problem that is solved using the shooting method. Here $N_T$ is the total mass enclosed by the boundary radius $r = r_f$, $N_T = N(r_f)$. Although solutions can be found for $m=0$ and $m=1$, in this study we {simplify our description and we decided} to work only with {the} case $m = 0$.

\begin{figure}[t!]
\centering
\includegraphics[width=0.45\textwidth]{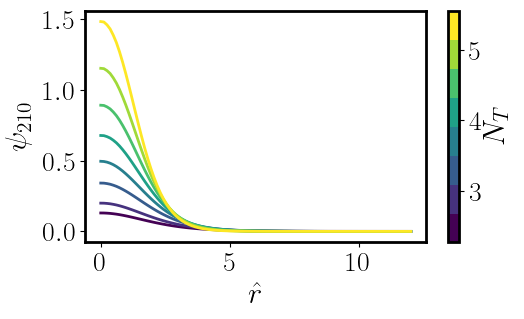}
\includegraphics[width=0.45\textwidth]{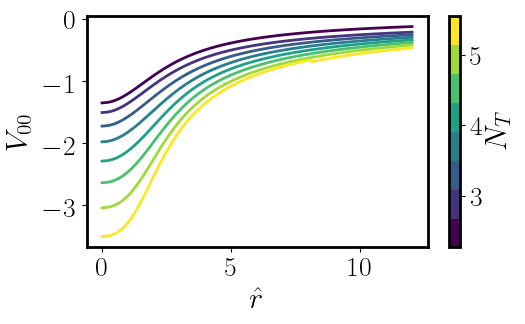}
\includegraphics[width=0.45\textwidth]{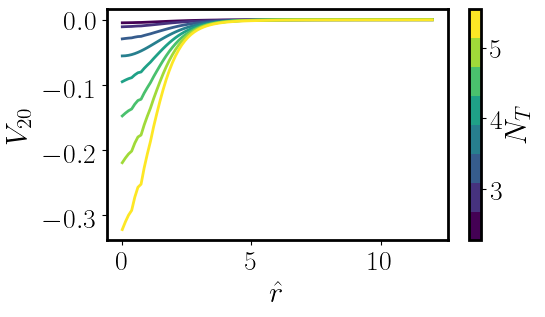}
\caption{Family of solutions of the multiSFDM $(\Psi_{100},\Psi_{210})$. The excited state radial function $\psi_{210}$ (upper), first function $V_{00}$ (middle panel) and second function $V_{20}$ (bottom panel) in the expansion of the potential $V$. In color scale, the total mass $N_T$ of each of the solutions in the family is shown.}
\label{fig:fammixstates210}
\end{figure}

We fix the central value $\psi_{100}(0)=1$ to find the eigen-values $E_{100}$ and $E_{210}$ and the initial values $V_{00}(0), V_{20}(0), \psi_{210}(0)$ of the bound multiSFDM configuration. We solve it in a fixed range of $(0, r_f)$ and we vary the boundary value $N_T$ to find a family of solutions. In Fig. \ref{fig:fammixstates210} we show the plots of $\psi_{210}, V_{00}$, and $V_{20}$ for the family of solutions we found.

{In Table \ref{tab:210fam} the different quantities that characterize each of the solutions of the family are shown: the total mass of the configuration $N_T$ (that we use as the solution identifier within the family); the energy eigen-values of the ground state $E_{100}$ and the excited state $E_{210}$; the total energy of the configuration $E_T= (E_{100}N_{100} + E_{210}N_{210})/N_T$; the mass ratio $\eta = N_{210}(r_f)/N_{100}(r_f)$ and amplitude ratio $\zeta = \psi_{100}(0)/\psi_{210}(0)$ between states of the configuration.}

\begin{deluxetable}{cccccc}
\tablecaption{multiSFDM $(100, 210)$. Total mass of the configuration (column 1), energy eigen-values of the ground (2) and excited state (3), total energy of the configuration (4), mass ratio between states of the configuration $\eta = N_{210}(r_f)/N_{100}(r_f)$ (5) and amplitude ratio between states of the configuration $\zeta = \psi_{100}(0)/\psi_{210}(0)$ (6). \label{tab:210fam}}
\tablewidth{0pt}
\tablehead{\colhead{$N_T$} & \colhead{$E_{100}$} &
\colhead{$E_{210}$}& \colhead{$E_T$} & \colhead{$\eta$} & \colhead{$\zeta$}}
\decimalcolnumbers
\startdata
2.1 & -0.69 & -0.40 & -0.69 & 0.01 & 37.27 \\
2.3 & -0.69 & -0.40 & -0.66 & 0.14 & 7.70  \\
2.5 & -0.84 & -0.54 & -0.77 & 0.29 & 5.01  \\
2.7 & -0.84 & -0.54 & -0.74 & 0.48 & 3.73  \\
3.0 & -1.03 & -0.72 & -0.90 & 0.71 & 2.93  \\
3.5 & -1.25 & -0.92 & -1.07 & 1.27 & 2.02  \\
4.0 & -1.51 & -1.16 & -1.28 & 1.97 & 1.47  \\
4.3 & -1.68 & -1.31 & -1.42 & 2.50 & 1.25  \\
4.5 & -1.80 & -1.42 & -1.52 & 2.90 & 1.12  \\
5.0 & -2.12 & -1.71 & -1.79 & 4.12 & 0.87  \\
5.5 & -2.49 & -2.04 & -2.11 & 5.83 & 0.67 \\ 
\enddata
\end{deluxetable}

\begin{figure*}[t!]
\centering
\includegraphics[width=0.4\textwidth]{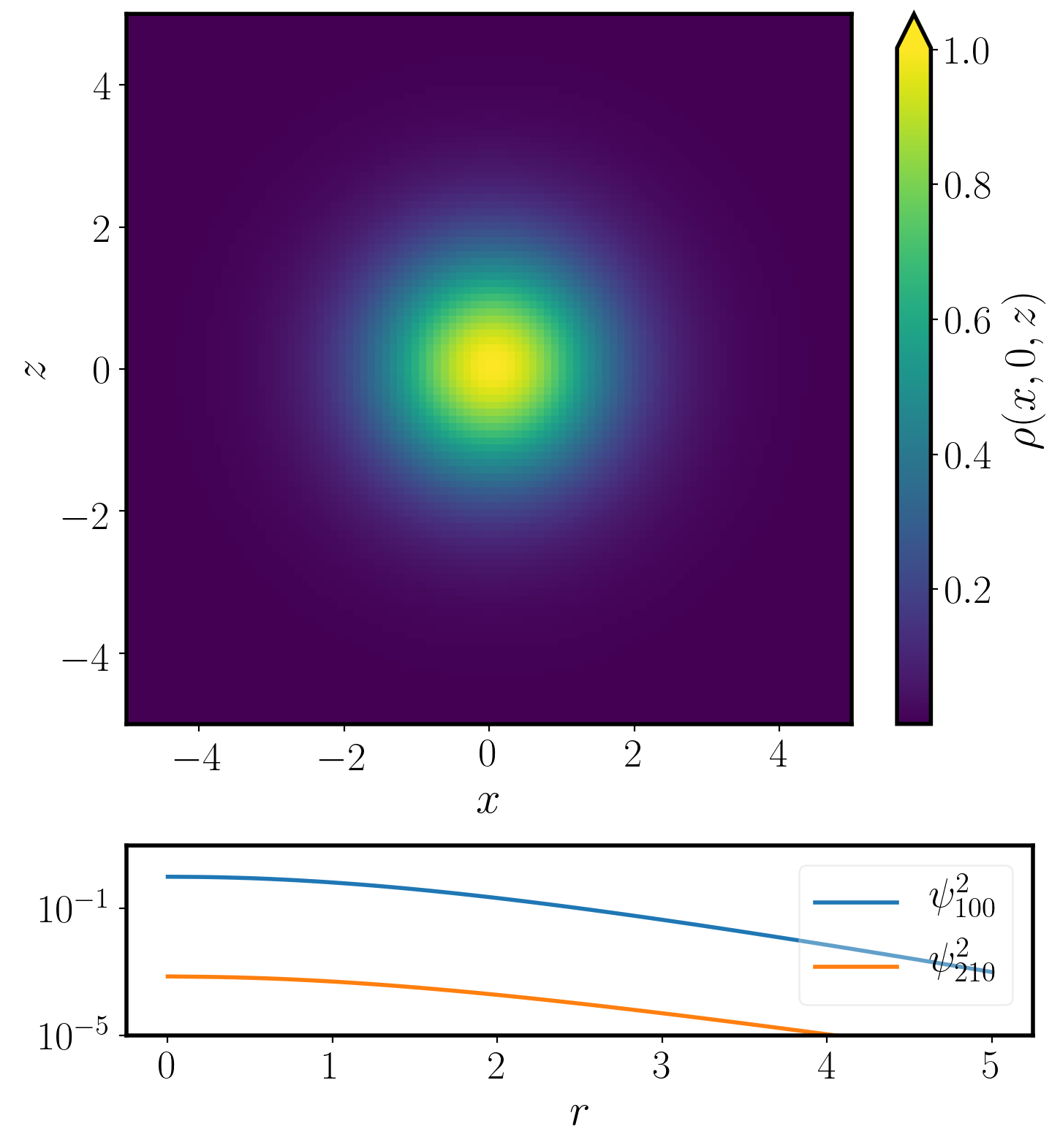}
\includegraphics[width=0.4\textwidth]{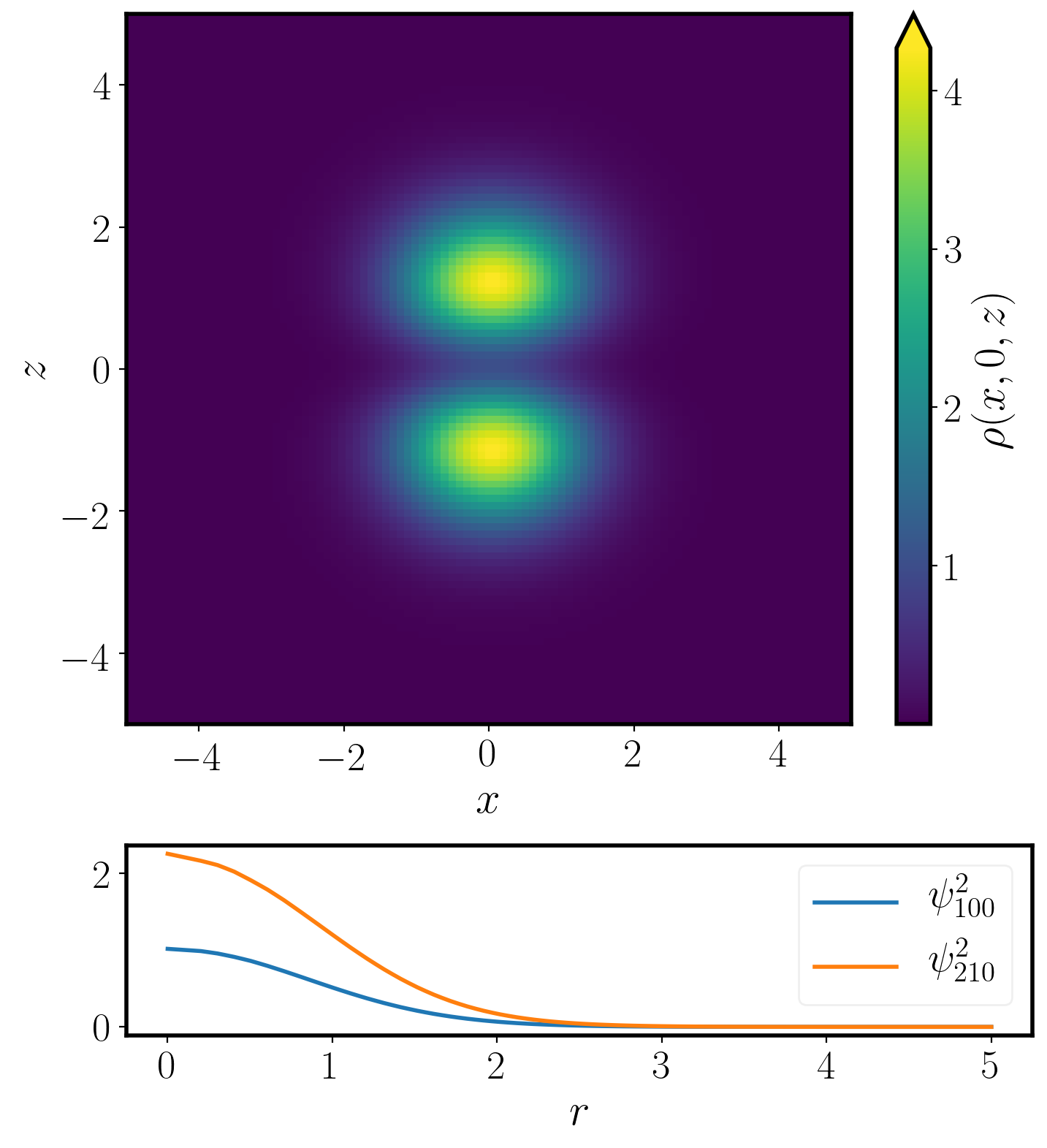}
\caption{Projection in the $(x,z)$ plane of the mass density as function of the ($x,y,z$) cartesian coordinates for the multiSFDM $(\Psi_{100},\Psi_{210})$. The left panel shows the solution with $N_T = 2.0$ where the monopole term $\psi_{100}$ dominates over the dipole term  $\psi_{210}$, the right panel is the solution with $N_T= 5.5$ where the excited state $\psi_{210}$ dominates. In color scale the mass density is shown.}
\label{fig:dens210rth}
\end{figure*}

In Figure \ref{fig:dens210rth} we show as representative examples two cases of the DM mass density $\rho = |\Phi_{100}|^2+|\Phi_{210}|^2$ as function of the $(r, \theta)$ coordinates, one solution with $N_T = 2.0$, where the monopole term $\psi_{100}$ dominates over the dipole term $\psi_{210}$, and the solution with $N_T= 5.5$ where the opposite happens.

\subsection{multiSFDM case (100,200)}
For the multiSFDM case $(\Psi_{100}, \Psi_{200})$ the system (\ref{eq:gpp2}) becomes
\begin{eqnarray} \label{eq:mix200}
&&\nabla^2_{r_0} \psi_{100}(r) = 2 (V_{00} - E_{100})\psi_{100}, \nonumber\\
&&\nabla^2_{r_0} \psi_{200}(r) = 2 (V_{00} - E_{200})\psi_{200}, \nonumber\\
&&\nabla^2_{r_0} V_{00}(r) = \psi_{100}^2 +  \psi_{200}^2, \nonumber
\end{eqnarray}
where the gravitational potential is simply 
\begin{equation*}\label{eq:pot200}
V (r, \theta) = \sqrt{4\pi} V_{00}(r) Y_{00}(\theta, \phi) = V_{00}(r)
\end{equation*}
and the circular velocity 
\begin{equation}\label{eq:halo3}
 v_h^2 = \frac{P_0}{r}.  
\end{equation}

\begin{figure}[t!]
\centering
\includegraphics[width=0.40\textwidth]{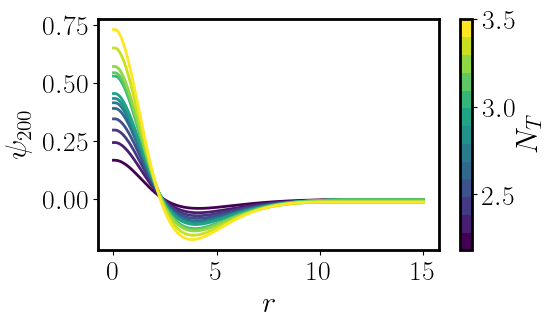}
\includegraphics[width=0.40\textwidth]{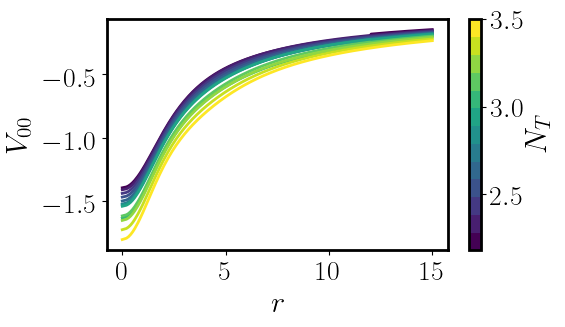}
\caption{Family of solutions of the multiSFDM $(\Psi_{100},\Psi_{200})$. The wave function $\psi_{200}$ (upper panel) and the potential $V$ (bottom panel). In color scale, the total mass $N_T$ of each of the solutions in the family is shown.}
\label{fig:fammixstates200}
\end{figure}

{In \cite{sph_multisates}, these multistate configurations were shown to be stable} only when $N_{200}(r_f)/N_{100}(r_f)<1.1$ so we restrict our selves to work only with this kind of solutions. Once again we use the total mass $N_T$ as a solution identifier  within the family. In Table \ref{tab:200fam} we show the energy eigen-values, the total energy, and the mass and amplitude ratios for each solution in the family. We also plot the corresponding family of solutions for this case in Fig. \ref{fig:fammixstates200}.

\begin{deluxetable}{cccccc}
\tablecaption{Same as in Table \ref{tab:210fam} but now using state $\psi_{200}$. \label{tab:200fam}}
\tablewidth{0pt}
\tablehead{\colhead{$N_T$} & \colhead{$E_{100}$} &
\colhead{$E_{200}$}& \colhead{$E_T$} & \colhead{$\eta$} & \colhead{$\zeta$}}
\decimalcolnumbers
\startdata
2.18  & -0.737    & -0.337    & -0.71 & 0.07   & 6.00    \\
2.30  & -0.745    & -0.341    & -0.70 & 0.14   & 4.11    \\
2.40  & -0.766    & -0.359    & -0.70 & 0.20   & 3.37    \\
2.50  & -0.788    & -0.377    & -0.70 & 0.27   & 2.90    \\
2.60  & -0.811    & -0.395    & -0.71 & 0.34   & 2.56    \\
2.66  & -0.830    & -0.412    & -0.72 & 0.38   & 2.41    \\
2.70  & -0.834    & -0.414    & -0.71 & 0.41   & 2.31    \\
2.75  & -0.840    & -0.418    & -0.71 & 0.45   & 2.20    \\
2.94  & -0.917    & -0.486    & -0.76 & 0.59   & 1.89    \\
2.97  & -0.896    & -0.463    & -0.73 & 0.62   & 1.84    \\
3.10  & -0.925    & -0.491    & -0.75 & 0.71   & 1.75    \\
3.30  & -0.977    & -0.532    & -0.77 & 0.88   & 1.54    \\
3.50  & -1.032    & -0.575    & -0.80 & 1.07   & 1.37 \\ 
\enddata
\end{deluxetable}

\section{CO-ADDED ROTATION CURVES: Data Analysis}

\subsection{URCs theory}

\begin{figure}[h!]
\centering
\includegraphics[width=0.40\textwidth]{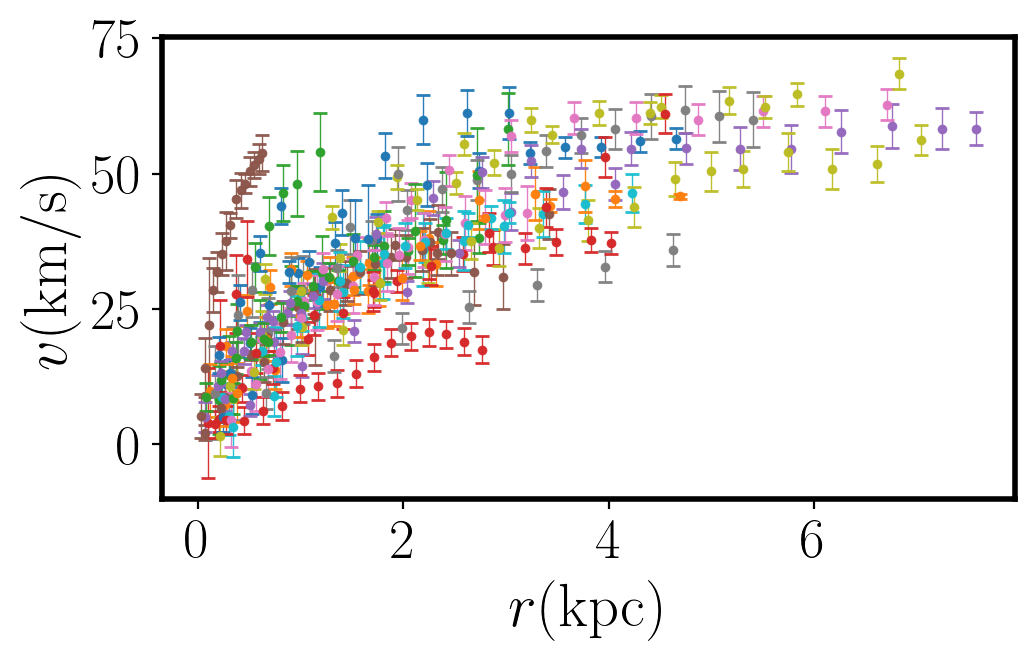}
\includegraphics[width=0.40\textwidth]{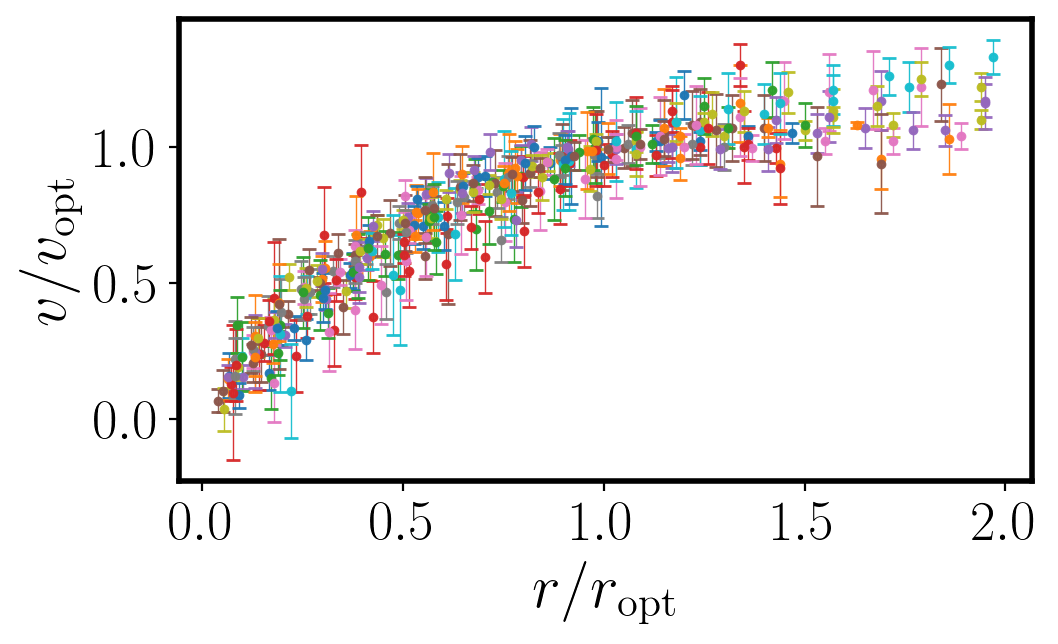}
\includegraphics[width=0.40\textwidth]{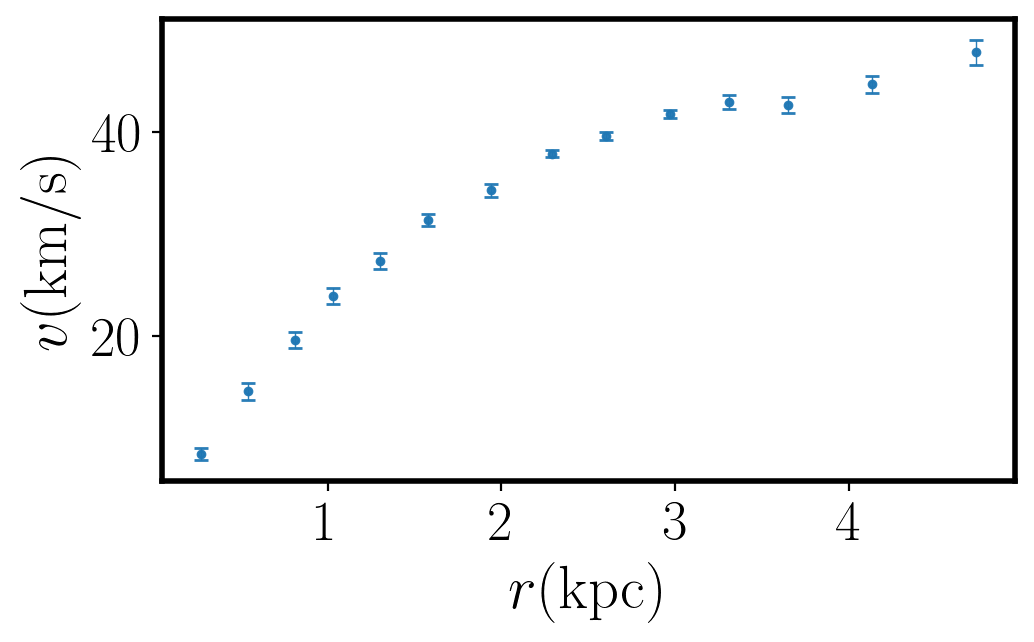}
\caption{Upper panel: Circular velocity measurements of 36 dwarf disc galaxies, middle panel: the normalized rotation curves and bottom panel: the co-added rotation curves of dwarf disc galaxies. Data from \cite{URCDD}.}
\label{fig:coadded}
\end{figure}

A co-added RC is a representative RC of a sample of galaxies with some particular properties in common (optical velocity $v_{opt}$, galaxy luminosity $L_B$ or absolute magnitude $M_K$). Once the radial coordinate and circular velocity measurements are normalized, all this RCs have the same shape and can be represented by only one co-added RC. In Fig. \ref{fig:coadded} we show the circular velocity measurements, the normalized data, and the co-added RC of the dwarf disc galaxies {as an example}.

The co-added RC is constructed first by setting a unique binning in the radial coordinate to all individual normalized RC. In each bin there should be only one velocity measurement (if more, then the velocities are averaged). Once this procedure has been done for all individual RCs, the next step is to compile all individual RCs into only one co-added RC, this is done by making a weighted average of all the velocity data in each bin.

The mass model of the co-added RC is called URC. After finding the best fitting parameters of the URC, we can apply the inverse transformation \cite[described in][]{URCDD} to find the best fitting parameters of each of the galaxies in the family.

\subsection{Mass models}

\subsubsection{Dwarf disc galaxies}
\label{sec:DD}

We use the co-added rotation curve from \cite{URCDD} that comes from a sample of 36 dwarf disc galaxies from the Local Volume catalog \citep{local_volume}. These galaxies have an exponential disk scale length $R_d$ in the range $(0.18, 1.63)\ \rm{kpc}$ and optical velocity $v_{\rm{opt}}=v(R_{\rm{opt}})$ in the range $(17,61)\ \rm{km/s}$. The optical radius $R_{\rm{opt}} = 3.2a_d$. Absolute magnitude $M_K \in (-19.9, -14.2)$.

To fit the co-added rotation curve of the dwarf spiral galaxies we use a simple model of a galaxy, consisting of a stellar disc, a HI disc, and a dark matter halo. The circular velocity of a particle due to these components is

\begin{equation*}
    v(r) = \sqrt{v_h^2 + v_d^2 + v_{HI}^2}
\end{equation*}

\noindent where $v_h, v_d$ and $v_{HI}$ are the circular velocities due to the halo and the stellar and HI discs, respectively.

The stellar disc is modeled using a razor-thin exponential disc profile whose surface mass density written in cylindrical coordinates $(\rho,\phi,z)$ is given by
\begin{equation*}
    \Sigma_d (\rho) = \Sigma_0 e^{-\rho/a_d},
\end{equation*}
where $a_d$ is the disc scale length,  $\Sigma_0$ is the central surface density and it is related to the total mass of the disc $M_d$ as $M_d = 2\pi\Sigma_0 a_d^2$. The circular velocity due to this density profile is \citep{freeman1970ApJ}
\begin{equation*}
    v_{d}(r) = \sqrt{\frac{2 G M_d y^2}{a} \left( I_0(y) K_0(y) - I_1 (y) K_1 (y) \right)},
\end{equation*}
where $I_n$ and $K_n$ are the modified Bessel functions of the first and second kind, respectively, and we have defined $y\equiv r/(2 a_d)$. 

The HI disc is also modeled using a razor-thin exponential disc profile but with $a_{HI} = R_{\rm{opt}}/3.2$, $R_{\rm{opt}}= 2.5\ \rm{kpc}$ and $M_{HI} = 1.7\times 10^{-8} M_\odot$.

\subsubsection{Low Surface Brightness galaxies}
\label{sec:LSB}
\begin{figure}
    \centering
    \includegraphics[width=0.45\textwidth]{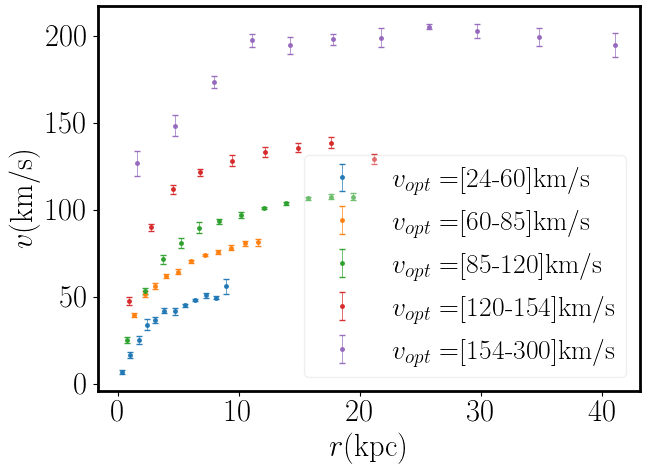}
    \caption{Co-added rotation curves for each of the five bins of the Low Surface Brightness galaxies. Data from \cite{URCLSB}.}
    \label{fig:URC_coadded}
\end{figure}

\cite{URCLSB} use a sample of 72 LSB galaxies with optical velocities in the range $v_{opt}\in (24,300)$ km/s and classify it in five groups (bins) depending on its optical velocity. 
Bin 1 with 13 galaxies, $v_{opt} \in (24, 60)$ km/s and mean disc scale length $a_d = 1.7$kpc.
Bin 2 with 17 galaxies, $v_{opt} \in (60, 85)$ km/s and mean disc scale length $a_d = 2.2$ kpc.
Bin 3 with 17 galaxies, $v_{opt} \in (85, 120)$ km/s and mean disc scale length $a_d = 3.7$ kpc.
Bin 4 with 15 galaxies, $v_{opt} \in (120, 154)$ km/s and mean disc scale length $a_d = 4.5$ kpc.
Bin 5 with 10 galaxies, $v_{opt} \in (154, 300)$ km/s and mean disc scale length $a_d = 7.9$ kpc.
When the individual  RC of the galaxies within a group are expressed in a normalized radius $r/R_{\rm{opt}}$ they all have almost the same distribution of matter. For each bin, \cite{URCLSB} calculated the co-added rotation curve that we will fit (see Fig. \ref{fig:URC_coadded}).

We model LSB galaxies with a stellar disc and a dark matter halo. The circular velocity of a particle due to these components is

\begin{equation*}
    v(r) = \sqrt{v_h^2 + v_d^2}
\end{equation*}

\noindent where $v_h$ and $v_d$ are the circular velocities due to the halo and the stellar disc, respectively. The stellar disc is modeled with the same exponential profile as the dwarf disc galaxies. For each co-added RC we use the mean disc scale length, so we end up with only one disc parameter $M_d$. 

In the case of bin 5, we also consider a galaxy bulge that is modeled using a velocity profile as suggested in \citep{URCLSB}:
\begin{equation*}
    v_b (r)= v_{in} \sqrt{\alpha \frac{r_{in}}{r}}.
\end{equation*}
where $r_{in} = 0.2 a_d$ is the radius of the innermost measure of the RC circular velocity $v_{in} = 127\  \rm{km/s}$, thus the only bulge parameter to fit is the $\alpha$ parameter.

\begin{figure}[t!]
\centering
\includegraphics[width=0.45\textwidth]{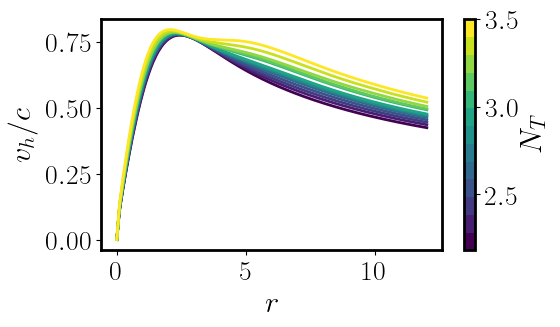}
\includegraphics[width=0.45\textwidth]{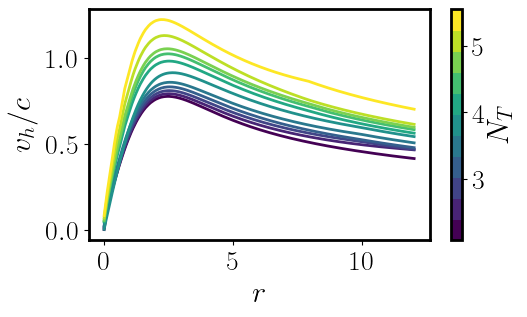}
\caption{Circular velocity $v_h/c$ for the $(\Psi_{100}, \Psi_{200})$ family (upper panel), the $(\Psi_{100}, \Psi_{210})$ family (bottom panel). In color scale, the total mass $N_T$ of each of the solutions in the family is shown.}
\label{fig:circvel}
\end{figure}

For the dark matter component we will use the circular velocity profiles (Equations \eqref{eq:halo} and \eqref{eq:halo3}) of all multistate configurations we have presented (see Fig. \ref{fig:circvel}). {Strictly speaking, our analysis should not be limited solely to the family of states that we have presented. However, exploring the entire parameter space of our system would result in a very large computational effort. For this reason, by restricting ourselves to this family of states, which cover different mass scales of the configurations quite well, we believe that it will be sufficient to give an estimate of the mass parameter of our model. In addition, with our results we can also put this model into context with the CDM model, which we will do later.}

Summarizing, we have a total of 3 parameters to fit, namely, $\sqrt{\lambda}$ (remember the {scaling} property described in Equation \eqref{rescaling}), $\tilde{\mu}$ and $M_d$, except for the case of bin 5 where we have an extra fitting parameter $\alpha$.

\subsection{Statistical calibration method}

We use the Markov Chain Monte Carlo (MCMC) method sampling the parameter space from uniform priors {(see Table \ref{tab:priors}).}

\begin{deluxetable*}{ccccccc}
\tablecaption{{Uniform priors used in the MCMC fitting.} \label{tab:priors}}
\tablewidth{0pt}
\tablehead{\colhead{Parameter} & \colhead{DD} & \colhead{} & \colhead{} & \colhead{LSB} & \colhead{} & \colhead{} \\
&& bin 1 & bin 2& bin3 & bin 4 &bin 5 }
\decimalcolnumbers
\startdata
$\sqrt{\lambda}$      & {$[10^{-7}, 1]$}  & {$[10^{-7}, 1]$} & {$[10^{-7}, 1]$}   & {$[10^{-7}, 1]$} & {$[10^{-7}, 1]$}  & {$[10^{-7}, 1]$}\\
$\mu$ ($\rm{eV}/c^2$) &  {$[10^{-26}, 10^{-18}]$} & {$[10^{-26}, 10^{-18}]$} & {$[10^{-26}, 10^{-18}]$} & {$[10^{-26}, 10^{-18}]$} &{$[10^{-26}, 10^{-18}]$} & {$[10^{-27}, 10^{-19}]$} \\
$M_d (10^{10} M_\odot )$ &  {$[10^{-6}, 10^{0}]$}  & {$[10^{-5}, 10^{1}]$}   & {$[10^{-6}, 10^1]$}         & {$[10^{-6}, 10^1]$}  & {$[10^{-5}, 10^2]$}  &  {$[10^{-5}, 10^2]$}  \\
$\alpha$ &&&&&& {$[10^{-6},10]$} \\
\enddata
\end{deluxetable*}
For each of the configurations of multiSFDM, we use $5 \times 10^4$ steps with $30$ \% burn-in and 50 walkers to sample the parameter space. The results for each one of the varied parameters were calculated using the Lmfit \citep{lmfit} and Emcee \citep{emcee} Python packages.
\section{Results and discussion}
\label{sec:discussion}

\subsection{Dwarf disc galaxies}
We performed the fit of the dwarf disc galaxies co-added RC with each of the solutions of both multiSFDM families. We select the best fit in each family using the Akaike information criterion (AIC) and the Bayesian information  criterion (BIC). {AIC gives a measure of the fit of a given model to the data. It measure the goodness of a fit and it gives a penalty on the number of parameters in the model. If the model is simpler (has few parameters) the penalty is less. The lower AIC value says that the model has better performance. The BIC works as the AIC but with a different penalty in the number of parameters in the model. In AIC, the penalty is $2k$, with $k$ being the number of parameters of the model, and in BIC the penalty is $\ln{(n)} k$, $n$ being the number of data points to fit.}
In Table \ref{tab:fitresDD} we present the results of the fit.

\begin{deluxetable*}{cccccccc}
\tablecaption{Fit results for the dwarf disc galaxies co-added rotation curve. Multi-SFDM family name (column 1), total mass of the configuration (2), reduced $\chi^2$ (3), the Akaike information criterion (4), the Bayesian information criterion (5), SFDM particle mass (6), scaling parameter (7), stellar disc mass (8). \label{tab:fitresDD}}
\tablewidth{0pt}
\tablehead{\colhead{Family} & \colhead{$N_T$} & \colhead{$\chi^2_{\rm{red}}$} & \colhead{AIC} & \colhead{BIC} & \colhead{$\mu \pm \sigma_{\mu}$} & \colhead{$\sqrt{\lambda} \pm \sigma_{\sqrt{\lambda}}$} & \colhead{$M_d \pm \sigma_{M_d}$} \\
&&&&& $(10^{-24}\rm{eV}/c^2)$ &$(\times 10^{-3})$ & $(10^{7}M_\odot)$}
\decimalcolnumbers
\startdata
$\Psi_{100}$ &  {1.5} & 1.6  & 9.5 & 11.4 & $17.4\pm 0.6$ &  {$0.191 \pm 0.001$}    & $12.00 \pm 1.37$ \\
$(\Psi_{100}, \Psi_{200})$ & 2.6 & 1.50 & 8.29 & 10.2 & $18.3 \pm 1.0$ & $0.183 \pm 0.002$ & $4.34 \pm 1.76$ \\
$(\Psi_{100}, \Psi_{210})$ &3.5 & 1.27 & 6.01 & 7.9 & $23.8 \pm 1.2$ & $0.154 \pm 0.002$ & $10.28 \pm 1.49$  \\
\enddata
\tablecomments{We only show the result of the best fit per family.}
\end{deluxetable*}

{To see whether the multistates give a better fit of the rotation curves, we also made the adjustment considering the dark matter halo in the ground state $\Psi_{100}$, {which is commonly used to describe the core in SFDM galaxies and it is also typically used to model dwarf-sized galaxies}. To do that, we use the Gaussian ansatz \citep{sc15,guzman2018head,mipaper1}:
\begin{equation}
    \rho(r)= \frac{M}{(\pi R_c^2)^{3/2}}e^{-r^2/R_c^2}
\end{equation}
as an approximation of the ground state density.} {We decided to use this Gaussian profile since {previous works \cite[see, for example,][]{mipaper1} have shown} that this profile can very well describe the numerical solution of the ground state configuration of the SP system. }

\begin{figure}
\centering
\includegraphics[width=0.43\textwidth]{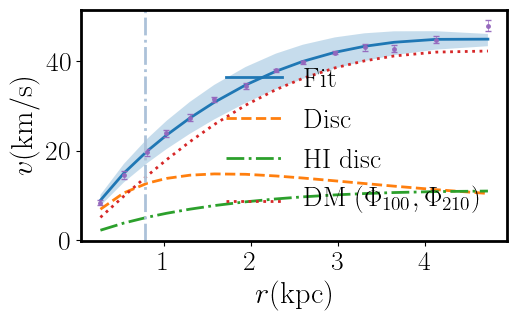}
\includegraphics[width=0.45\textwidth]{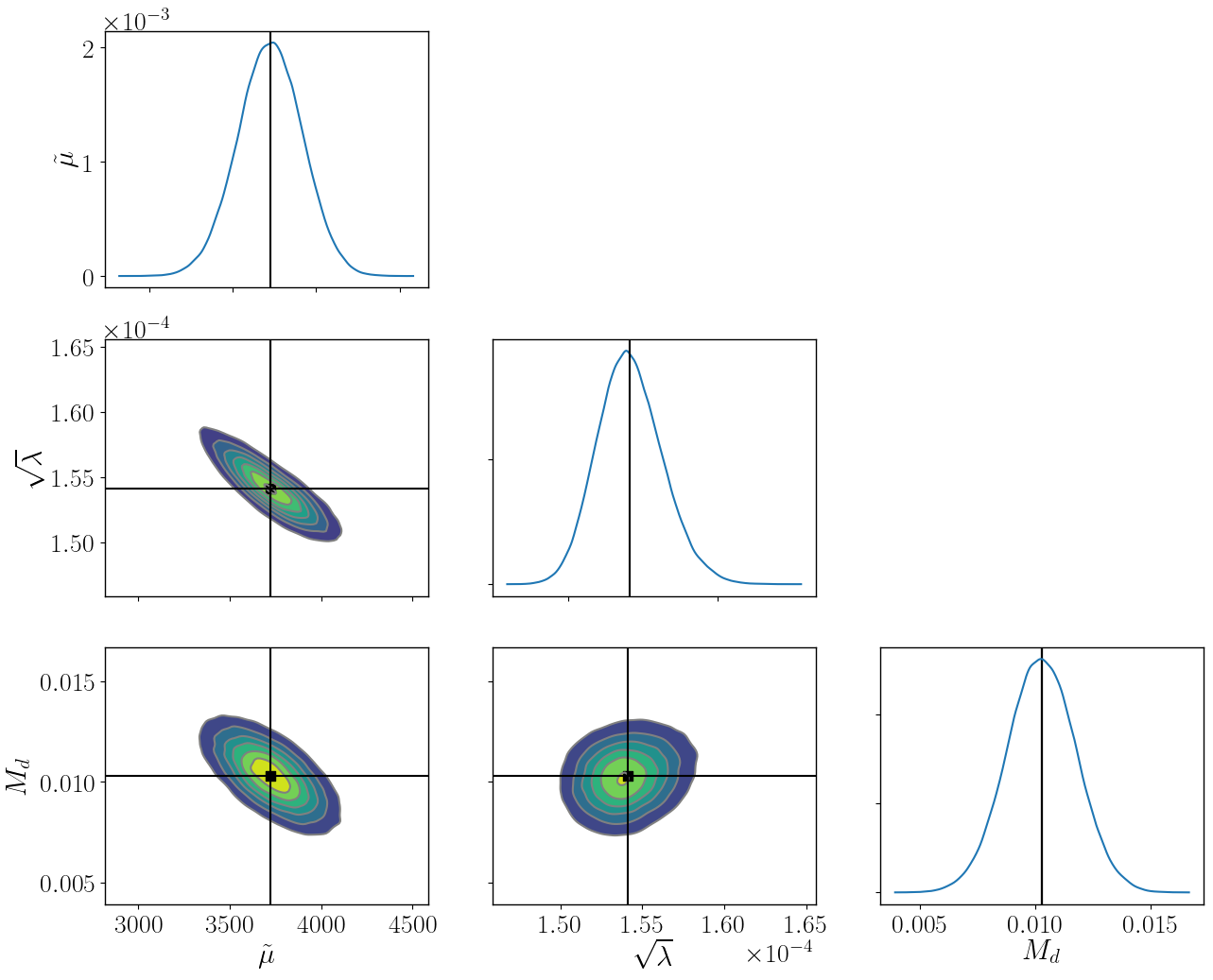}
\caption{{Upper panel: }Dwarf Disc galaxies co-added rotation curve. {The disc, HI disc, and DM contributions are also shown}. Dark matter is in the multiSFDM $(\Psi_{100}, \Psi_{210})$. The best fit parameters are shown in Table \ref{tab:fitresDD}. The horizontal line is the disc characteristic length $a_d$. {Bottom panel: We show the posterior distribution of parameters, as an example, in this case. Particle mass $\tilde{\mu}$ is in 1/kpc units and disc mass $M_d$ is in $10^{10}M_\odot$ units.}}
\label{fig:DDfit}
\end{figure}

If we use the AIC, BIC, and $\chi^2_{\rm{red}}$ we can state that the best fit is obtained with the multiSFDM $(\Psi_{100}, \Psi_{210})$ model, particularly the solution characterized with the total mass $N_T = 3.5$ and having a particle mass $\mu = (2.38 \pm 0.12) \times 10^{-23} \rm{eV}/c^2$. In {the upper panel of} Fig. \ref{fig:DDfit} we show the plot of the fit and the contribution of the disc, HI disc, and DM separately; {and in the bottom panel we show a corner plot of the posterior distribution of the fitting parameters}.

The fit with this DM model is consistent with having a stellar disc mass $M_d \approx 10^{8} M_\odot$, that is also consistent with the one obtained in \cite{URCDD} with the Burkert profile as DM model.

\subsection{LSB galaxies}
As in the case of the DD galaxies, we also performed the fit of the five different bins of the LSB galaxies co-added RC with each of the solutions of both multiSFDM families, we select the best fit in each family using the AIC and BIC parameters and in Table \ref{tab:fitresLSB} we present the results of the best fit found for each of the two families of configurations for all five bins. For bins 1,2, and 3 we also show the best fit using only the ground state as the DM halo with the Gaussian ansatz. For bins 4 and 5, it is not possible to fit the co-added RC using a single state.  This is expected since these bins are where the largest and most massive galaxies belong, so it would be expected that the base state alone would not be able to model these galaxies.

\begin{deluxetable*}{cccccccccc}
\tablecaption{Fit results for the LSB galaxies co-added rotation curves. LSB bin number (column 1), multi-SFDM family name (2), total mass of the configuration (3), reduced $\chi^2$ (4), the Akaike information criterion (5), the Bayesian information criterion (6), SFDM particle mass (7), scaling parameter (8), stellar disc mass (9) and bulge parameter (10). \label{tab:fitresLSB}}
\tablewidth{0pt}
\tablehead{\colhead{Bin} &\colhead{Family} & \colhead{$N_T$} & \colhead{$\chi^2_{\rm{red}}$} & \colhead{AIC} & \colhead{BIC} & \colhead{$\mu \pm \sigma_{\mu}$} & \colhead{$\sqrt{\lambda} \pm \sigma_{\sqrt{\lambda}}$} & \colhead{$M_d \pm \sigma_{M_d}$} & \colhead{$\alpha \pm \sigma_\alpha$} \\
&&&&& & $(10^{-24}\rm{eV}/c^2)$ &$(\times 10^{-3})$ & $(10^{7}M_\odot)$&}
\decimalcolnumbers
\startdata
    & $\Psi_{100}$  &   {1.5}    & 1.3  & 6.1 & 7.6 & $7.3\pm 0.7$ &  {$0.214 \pm 0.003$}     & $63.5 \pm 11.9$ &\\
1   & $(\Psi_{100}, \Psi_{200})$ & 3.10 & 1.44 & 6.900 & 8.3 & $7.4 \pm 1.02$ & $0.205 \pm 0.005$ & $37.5 \pm 16.0$  &\\
    & $(\Psi_{100}, \Psi_{210})$ & 3.5   & 1.29 & 5.624 & 7.1 & $10.5 \pm 1.40$ & $0.173 \pm 0.005$ & $55.0 \pm 13.5$ & \\ \hline
    & $\Psi_{100}$&  {1.3}   & 1.2  & 4.1 & 5.3 & $2.1 \pm 0.4$ &    {$0.374 \pm 0.021$}  &$411.7 \pm 19.9$&  \\
2   & $(\Psi_{100}, \Psi_{200})$ & 2.94  & 1.02           & 2.677   & 3.9   & $2.0 \pm 0.47$  & $0.350 \pm 0.030$ & $362.5 \pm 20.0$    &               \\
    & $(\Psi_{100}, \Psi_{210})$ & 5.5   & 0.91           & 1.496   & 2.7   & $3.4 \pm 0.55$  & $0.221 \pm 0.014$ & $354.9 \pm 19.3$    &               \\ \hline
    & $\Psi_{100}$& {0.8} & 0.2  & -17.4 & -16.0 & $1.20 \pm 0.13$ & {$0.446 \pm 0.022$} & $1381.5 \pm 66.4$&                      \\
3   & $(\Psi_{100}, \Psi_{200})$ & 3.50 & 0.19 & -17.211 & -15.8 & $1.1 \pm 0.16$ & $0.419 \pm 0.020$ & $1238.0 \pm 70.8$  &               \\
    & $(\Psi_{100}, \Psi_{210})$ & 3.5 & 0.19 & -17.299 & -15.8 & $1.6 \pm 0.23$ & $0.361 \pm 0.017$ & $1349.8 \pm 66.6$ &               \\ \hline
4   & $(\Psi_{100}, \Psi_{200})$ & 2.18  & 5.24           & 17.255  & 17.8  & $1.3 \pm 0.26$  & $0.373 \pm 0.027$ & $4295.4 \pm 131.0$  &               \\
    & $(\Psi_{100}, \Psi_{210})$ & 5.5   & 4.77           & 16.410  & 17.0  & $1.4 \pm 0.40$  & $0.262 \pm 0.026$ & $4280.4 \pm 128.9$  &               \\ \hline
5   & $(\Psi_{100}, \Psi_{200})$ & 2.50  & 1.73           & 9.086   & 10.7  & $0.24 \pm 0.16$ & $0.716 \pm 0.196$ & $16806.5 \pm 576.9$ & $0.8 \pm 0.1$ \\
    & $(\Psi_{100}, \Psi_{210})$ & 4.0   & 1.73           & 9.076   & 10.7  & $0.39 \pm 0.21$ & $0.525 \pm 0.117$ & $16793.1 \pm 583.1$ & $0.8 \pm 0.1$        \\
\enddata
\tablecomments{We only show the result of the best fit per family.}
\end{deluxetable*}

\begin{figure}[!ht]
\centering
\includegraphics[width=0.43\textwidth]{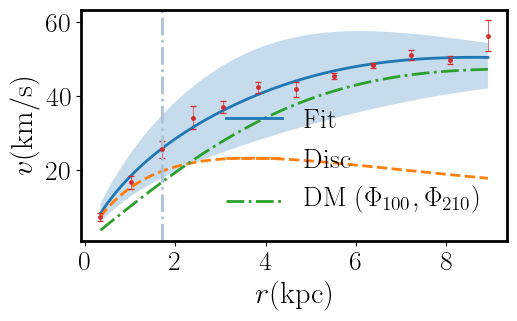}
\caption{{LSB bin 1 galaxies co-added rotation curve. {The disc and DM contributions are also shown}. Dark matter is in the multiSFDM $(\Psi_{100}, \Psi_{210})$. The best fit parameters are shown in Table \ref{tab:fitresLSB}. The horizontal line is the disc characteristic length $a_d$.}}
\label{fig:bin1fit}
\end{figure}
\begin{figure}[!ht]
\centering
\includegraphics[width=0.43\textwidth]{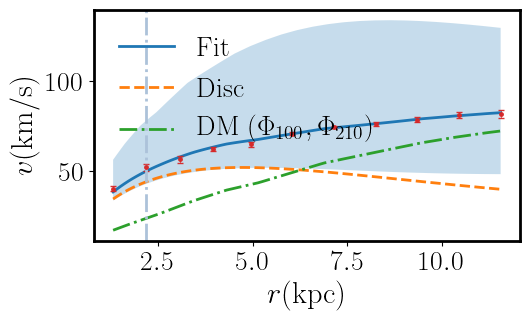}
\caption{Same as in Fig. \ref{fig:bin1fit} but for the LSB bin 2.}
\label{fig:bin2fit}
\end{figure}
\begin{figure}[!ht]
\centering
\includegraphics[width=0.43\textwidth]{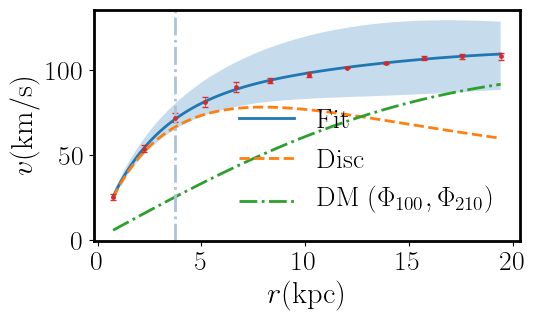}
\caption{Same as in Fig. \ref{fig:bin1fit} but for the LSB bin 3.}
\label{fig:bin3fit}
\end{figure}

\begin{figure}[!ht]
\centering
\includegraphics[width=0.43\textwidth]{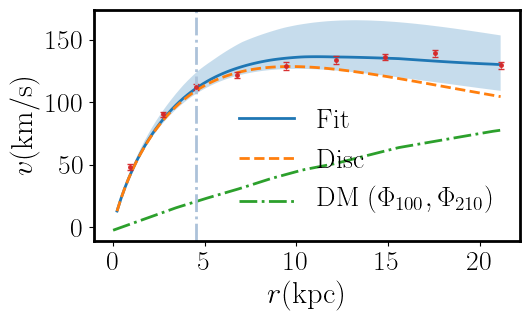}
\caption{Same as in Fig. \ref{fig:bin2fit} but for the LSB bin4.}
\label{fig:bin4fit}
\end{figure}

\begin{figure}[!ht]
\centering
\includegraphics[width=0.43\textwidth]{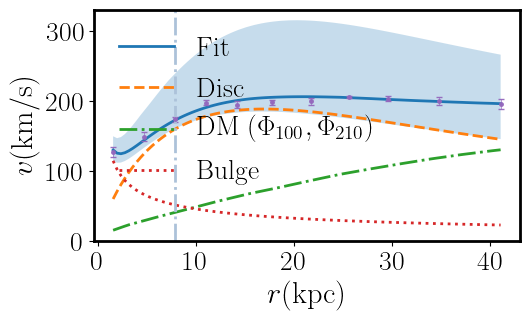}
\caption{{LSB bin 5 galaxies universal rotation curve. {The disc, bulge, and DM contributions are also shown.} Dark matter is in the multiSFDM $(\Psi_{100}, \Psi_{210})$. The best fit parameters are shown in Table \ref{tab:fitresLSB}. The horizontal line is the disc characteristic length $a_d$.}}
\label{fig:bin5fit}
\end{figure}

For LSB bins 1 2, 4 and 5 the best model of dark matter turns out to be again, multiSFDM $(\Psi_{100}, \Psi_{210})$ while for bin 3 the best model is the ground state $\Psi_{100}$.

For LSB bins 1,2, 3, and 4, the particle mass $\mu$ of order $10^{-24} \rm{eV}/c^2$ is prefered, however for the largest LSB galaxies (LSB bin 5) the particle mass is smaller $\mu = (3.9 \pm 2.1) \times 10^{-25} \rm{eV}/c^2$, the same order of magnitude that spiral galaxies like the Milky Way have \citep{satellites}. However, it would be expected that the reason why these lighter masses are preferred in this bin is because the largest and most massive galaxies belong to it, so configurations with only one excited state should not really describe this type of galaxies correctly.

\begin{figure}
    \centering
    \includegraphics[width=0.4\textwidth]{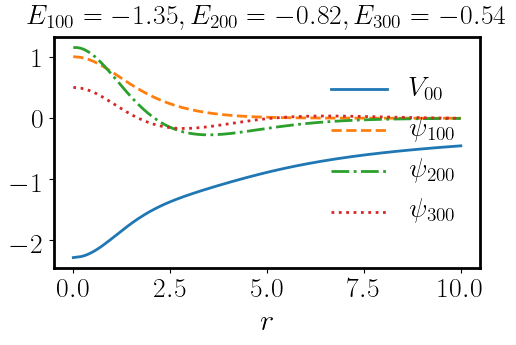}
    \caption{Three-states $(\Psi_{100}, \Psi_{200}, \Psi_{300})$ multistate configuration. {The three wave functions and the gravitational potential are shown}.}
    \label{fig:3states}
\end{figure}

For this particular bin, let us consider a three-state spherically symmetric multistate configuration made of the first three spherical states $(\Psi_{100}, \Psi_{200}, \Psi_{300})$, with energy eigenvalues $E_{100}= -1.35$, $E_{200}= -0.82$ and $E_{300}= -0.54$, the configuration have a total mass $N_T = 4.51$. In Fig. \ref{fig:3states} we show the plot of the solution, the three wave functions $\psi_{100}$ (zero nodes) $\psi_{200}$ (one node), $\psi_{300}$ (two nodes) and the potential $V$.

\begin{figure}
    \centering
    \includegraphics[width=0.4\textwidth]{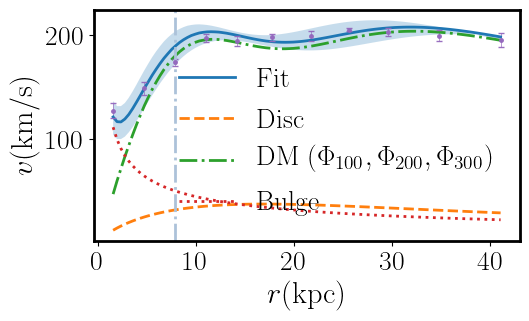}
    \caption{LSB bin 5 fit with a three-states $(\Psi_{100}, \Psi_{200}, \Psi_{300})$ spherically symmetric multistate configuration. The scalar DM mass $\mu = (1.24 \pm 0.06) \times 10^{-24} \rm{eV}/c^2$ becomes bigger than for a two-states configuration.}
    \label{fig:3statesfit}
\end{figure}

 In Fig. \ref{fig:3statesfit} we show the fit of the LSB bin 5 co-added RC, the mass of the multiSFDM $\mu = (1.24 \pm 0.06) \times 10^{-24} \rm{eV}/c^2$ becomes bigger than for a two-state configuration. The rest of the fit parameters take the values $\sqrt{\lambda} = (0.777 \pm 0.009) \times 10^{-3}$, $M_d= (646.8 \pm 315.3) \times 10^7 M_\odot$ and $\alpha = 0.8 \pm 0.1$. We note that this configuration beside that it allows a bigger DM particle mass, it has the oscillations seen in the data.  

\subsection{NFW}

From $N-$body simulations of CDM, \cite{NFW} found an equilibrium density profile for DM halos
\begin{equation*}
    \rho(r)= \frac{\rho_{0}}{\left(r/r_{s}\right) \left(1 + r/r_{s} \right)^2}
\end{equation*}
where $r_s$ is the scale radius and $\rho_0$ is a characteristic density. 

The halo circular velocity contribution is
\begin{equation}
    v_h = \sqrt{\frac{GM(r)}{r}},   
\end{equation}
{where $M(r)$ is the enclosed mas at radius $r$ given by
\begin{equation}
    M(r) = 4\pi r_s^3 \rho_0\left(-\frac{r}{r + r_s} + \ln \left( \frac{r + r_s}{r} \right) \right),
\end{equation}
which give us a two ($r_s, \rho_0$) parameters profile.
}

We performed the same MCMC fitting procedure we did with the SFDM model. In Table \ref{tab:fitresNFW} we show the results for the LSB bins 1 to 5, the reduced $\chi^2$, the AIC and BIC criteria, and the best fitting parameters. For the case of DD galaxies, we could not fit the rotation curves with the NFW profile, which is consistent with the fact that DD galaxies necessarily need a core to be able to explain their rotation curves.

Comparing the AIC, BIC, and $\chi^2$ we see that multiSFDM can better describe the LSB co-added RC for bins 1, 2, 3, and 5 than the NFW profile. This turns out to be very interesting, since our model, as simple as it seems in only adopting {two-state} configurations, seems to fit the data better than the standard cosmological model. It is clear that if we continue to increase {the number of states}, we will adjust the rotation curves better and better, which will reduce the $\chi^2$ of our model, although this will also result in a greater penalty for the model. In this way, we would expect there to be a preferred number of states where the value of our selection criteria (AIC and BIC) would be reduced to the {minimum}, even less than those reported by our model with only two states. Thus, we would expect that, in general, the multiSFDM model would be preferred for universal rotation curves than CDM.

The disc mass parameter $M_d$ disagree with the results of \cite{URCLSB}, with the NFW profile the expected stellar disc mass is smaller.

\begin{deluxetable*}{ccccccc}
\tablecaption{LSB rotation curves fitting results with a NFW profile. Velocity bin (column 1), reduced $\chi^2$ (2), the Akaike information criterion (3), the Bayesian information criterion (4), stellar disc mass (5),  scale radius (6) and characteristic density (7). \label{tab:fitresNFW}}
\tablewidth{0pt}
\tablehead{\colhead{bin} & \colhead{$\chi^2_{red}$} & \colhead{AIC}   & \colhead{BIC}  & \colhead{$M_d \pm \sigma_{M_d}$} & \colhead{$r_s \pm \sigma_{r_s}$} & \colhead{$\rho_0 \pm \sigma_{\rho_0}$} \\
    &   &    &    & $(10^{7}M_\odot)$ & (kpc) & $(10^{-4}M_\odot /\rm{pc}^3)$}
\decimalcolnumbers
\startdata
1   & 4.23         & 19.84 & 21.3 & $19.2 \pm 10.88$        & $64.4 \pm 23.1$         & $2.156 \pm 1.076$           \\
2   & 1.72         & 4.91 & 6.4  & $168.5 \pm 47.91$        & $34.8 \pm 23.1$         & $8.235 \pm 5.979$           \\
3   & 1.14         & 4.16  & 5.6  & $873.7 \pm 73.70$        & $109.1 \pm 43.4$         & $2.162 \pm 1.004$           \\
4   & 1.51         & 6.06  & 6.7  & $2144.1 \pm 465.47$       & $14.2 \pm 2.7$          & $52.398 \pm 20.356$          \\
5   & 1.78         & 9.39  & 11.0 & $12891.1 \pm 1229.26$      & $47.6 \pm 20.6$         & $9.275 \pm 6.097$           \\
\enddata
\end{deluxetable*}
\section{Conclusions}
\label{sec:concl}
{In this work we consider spherically symmetric and axi-symmetric multistate scalar dark matter as dark matter halos in dwarf disc and low surface brightness galaxies. The multistate configurations are equilibrium solutions of the Gross-Pitaevskii-Poisson equations when the boson particles are in more than one state. Particularly, we work in multistate configurations where bosonic particles are able to be in the ground state ($\Psi_{100}$) and one excited state ($\Psi_{210} $ or $\Psi_{200}$).}

{We test this model by fitting co-added rotation curves of LSB galaxies and dwarf disc galaxies using a MCMC method. We determine the parameters that provide the best fit to data. The resulting parameters of the baryonic mass model are consistent with the ones found in similar works that use different DM models \citep{URCDD, URCLSB}}

Both in LSB galaxies as in dwarf disc galaxies, the multistates models fit better the rotation curves than a single ground state. {In LSB bins 1,2,3, and 5 the multistates models fit better the rotation curves than the NFW profile, only in LSB bin 4 galaxies NFW profile describes better the rotation curve.}
Multi states are promising candidates to scalar field dark matter halos in large galaxies since the particle mass needed for them ($\mu \sim (10^{-23} - 10^{-24})\rm{eV}/c^2$) is larger than the required if the halo is composed only with the ground state $\mu \leq 10^{-25}\rm{eV}/c^2$. It suggest that adding more or higher excited states in large halos will increase the mass to $\mu \sim  (10^{-22} - 10^{-23})\rm{eV}/c^2$ to be in agreement with dSph galaxies and cosmological constraints of the scalar field particle mass. {The addition of excited states postpones the Newtonian drop in the circular velocity to greater distances which makes it have a smaller extension and therefore a greater particle mass.}

These results encourage further studies on different configurations of multistates scalar field dark matter halos.

\vspace{5mm}

\software{Lmfit Python package \citep{lmfit}, Emcee Python package \citep{emcee}}
          
\acknowledgments
This work was partially supported by CONACyT M\'exico under grants: A1-S-8742, 304001, 376127; Project No. 269652 and Fronteras Project 281; Xiuhcoatl and Abacus clusters at Cinvestav, IPN; I0101/131/07 C-234/07 of the Instituto Avanzado de Cosmolog\'ia (IAC) collaboration (http:// www.iac.edu.mx). J.S. acknowledges financial support from a CONACyT doctoral fellowship. L.P. acknowledges sponsorship from CONACyT through grant CB-2016-282569.

\bibliography{references.bib}{}
\bibliographystyle{apalike}
\end{document}